\documentclass[12pt]{article}
\def\lae{\;^{<}_{\sim} \;} \def\gae{\; ^{>}_{\sim} \;}

\title{\textbf{Super Hilltop Inflation}}
{\author{\\[1cm]
{\sc \large Chia-Min Lin$^{1,\ast}$ and Kingman Cheung$^{1,2,3,\dag}$}\\
{\sl\small $^1$Department of Physics, National Tsing Hua University, Hsinchu, Taiwan 300 }\\
{\sl\small $^2$Physics Division, National Center for Theoretical Sciences,
Hsinchu 300, Taiwan}\\
{\sl\small $^3$Division of Quantum Phases \& Devices, School of Physics,
Konkuk university, Seoul 143-701, Korea}
}}

\usepackage[margin=2cm]{geometry}
\usepackage{graphicx,color}
\usepackage{subfigure}
\begin{document}
\maketitle
\begin{abstract}
  In this paper, we consider logarithmic radiative corrections and
  higher order terms to the supersymmetric hilltop F- and D-term
  hybrid inflation models. Conventional F- and D-term hybrid
  inflation only predicts $n_s \gae 0.98$. We show that via a positive
  quadratic and a negative quartic correction the spectral index can
  be reduced to $n_s=0.96$ suggested from latest WMAP result and also
  cosmic string problem appeared in SUSY hybrid inflation can be
  solved with mild tuning of the parameters if $\kappa \lae 0.01$ for
  F-term inflation and $g \lae 0.05$ for D-term inflation.
\end{abstract}
\footnoterule{\small $^\ast$cmlin@phys.nthu.edu.tw,
$^\dag$cheung@phys.nthu.edu.tw}
\section{Introduction}
\label{1} Inflation \cite{Starobinsky:1980te, Sato:1980yn,
Guth:1980zm} (for review, \cite{Lyth:1998xn, Lyth:2007qh,
Linde:2007fr}) is an vacuum dominated epoch in the early universe
when the scale factor grew exponentially. This scenario is used to
set the initial condition for the hot big bang and provided
primordial density perturbation as the seed of structure formation.
In the framework of slow-roll inflation, the slow-roll parameters
are defined by
\begin{equation}
\epsilon \equiv \frac{M_P^2}{2} \left(\frac{V^\prime}{V}\right)^2,
\end{equation}
\begin{equation}
\eta \equiv M_P^2\frac{V^{\prime\prime}}{V},
\end{equation}
where $M_P=2.4\times 10^{18} \mbox{ GeV}$ is the reduced Planck mass.
The spectral index can be expressed in terms of
 the slow-roll parameters as
\begin{equation}
n_s=1+2\eta-6\epsilon.
\label{n}
\end{equation}
The spectrum is given by
\begin{equation}
P_R=\frac{1}{12\pi^2M_P^6}\frac{V^3}{V'^2}
\end{equation}
With the slow-roll approximation
the value of the inflaton field $\phi$,
in order to achieve $N$ e-folds inflation, is
\begin{equation}
N=M^{-2}_P\int^{\phi(N)}_{\phi_{end}}\frac{V}{V'}d\phi.
\label{efolds}
\end{equation}
From observation \cite{Komatsu:2008hk}, $P^{1/2}_R\simeq 5\times 10^{-5}$ at $N \simeq 60$.

Suppose we are going to build a small field inflation model, i.e.
$\phi \lae M_P$. For a wide range of the potentials, for example,
polynomial and logarithmic forms, we have
\begin{equation}
V^\prime \simeq \phi V^{\prime\prime}.
\end{equation}
Therefore, since $\phi \lae M_P$
\begin{equation}
|\epsilon|=\frac{\phi^2}{2M_P^2}|\eta|^2 \ll |\eta|.
\end{equation}
From Eq. (\ref{n})
\begin{equation}
n_s \simeq 1+2\eta.
\end{equation}
The latest WMAP 5-year result prefers the spectral index around
$n_s=0.96$ \cite{Komatsu:2008hk} which implies $\eta=-0.02$. This
suggests the inflation took place near a maximum of the potential
which is concave-downward. This kind of models is called ``hilltop
inflation'' \cite{Boubekeur:2005zm, Kohri:2007gq}.

In hybrid inflation \cite{Linde:1993cn}, a false vacuum is provided
by the waterfall field. By adding the false vacuum, even use a
quadratic potential (like the one in chaotic inflation
\cite{Linde:1983gd}), the model can still be a small field model
\cite{Copeland:1994vg}. 
In the framework of supersymmetry, there are two standard types of hybrid
inflation models: F-term \cite{Dvali:1994ms,
  Copeland:1994vg} and D-term  \cite{Binetruy:1996xj,
  Halyo:1996pp, Riotto:1997wy, Lyth:1997pf}. Both types of models
predict the spectral index $n_s \gae 0.98$. In conventional forms of
models, cosmic strings are produced after inflation. However, there
is a $10\%$ upper bound for the contribution from cosmic strings to
the CMB angular power spectrum \cite{Pogosian:2003mz,
Pogosian:2004ny, Wyman:2005tu}. This puts a very strong constraint
to the parameters in both F- and D-term hybrid inflation. In
\cite{Lin:2006xta}, it was shown that the cosmic string problem for
D-term inflation can be solved by using a negative quadratic
correction to the scalar potential, but a spectral index $n_s<0.96$
is required. In D-term inflation, if we reduce the superpotential
coupling to $\lambda \lae O(10^{-4}-10^{-5})$ and the $U(1)_F$ gauge
coupling $g \lae 2\times 10^{-2}$ \cite{Endo:2003fr, Rocher:2004my},
then the cosmic string contribution to the CMB can be reduced to
less than $10\%$. This also makes $n_s \simeq 1$. For $10\%$ cosmic
string contribution, $n_s \simeq 1$ is actually what we would like
to have \cite{Bevis:2007gh, Battye:2006pk}. However if the cosmic
string contribution is further reduced, for example, less than
$5\%$, $n_s \simeq 1$ is not favored. On the other hand, in the
F-term inflation, the constraint for the superpotential coupling
from cosmic string study is $\kappa \lae 7\times 10^{-7}$
\cite{Rocher:2004my, Rocher:2004et}. The above parameters will be
defined in the subsequent sections.

It is well-known that there is the $\eta$-problem \cite{Copeland:1994vg,
  Dine:1995uk} in the F-term hybrid inflation. But if we tune the coupling
mildly, it can be used as '$\eta$-correction'. If the correction is
negative \cite{BasteroGil:2006cm}, $n_s=0.96$ can be achieved. In this
paper, we show that we can also have $n_s=0.96$ in the case the
quadratic correction is positive if we include a negative quartic
term. Similar models can be realized in D-term hybrid inflation, where
the higher order correction terms can come from non-minimal gauge
kinetic function.

It would be interesting if we can have \emph{both} $n_s=0.96$ and solving the cosmic string problem. In this paper, we show that by considering some generic higher order terms, the spectral index can be reduced to fit the data and the cosmic string problem can also be solved.

This paper is organized as follows. In section \ref{2}, we introduce
a simple parameterization of our model and show the analytic
solutions of it. In section \ref{cosmicstring}, the formation of
cosmic string and its contribution to the CMB anisotropy is
explained. In section \ref{3}, we discuss the hilltop version of
D-term hybrid inflation. In section \ref{4}, we discuss the F-term
hybrid inflation case. Finally, we present our conclusions.

\section{The Potential}
\label{2}
\begin{figure}[htbp]
\begin{center}
\includegraphics[width=0.45\textwidth]{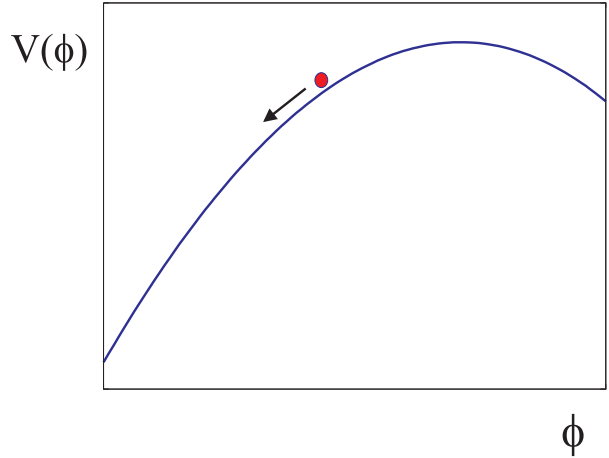}
\caption{The typical form of
$V(\phi)=V_0\left(1+\frac{a}{2}\phi^2-\frac{b}{4}\phi^4+
c\ln\left(\frac{\phi}{\Lambda}\right)\right)$ for
$a, b, c>0 $.
The red spot represents the slow-rolling inflaton field.}
\label{sketch}
\end{center}
\end{figure}

In this section, for simplicity, we set the reduced Planck mass
$M_P=1$. Consider the following form of the scalar potential
\begin{equation}
V(\phi)=V_0\left(1+\frac{a}{2}\phi^2-\frac{b}{4}\phi^4+c\ln\left(\frac{\phi}{\Lambda}\right)\right)
\label{potential}
\end{equation}
where $a,b,c>0$. A sketch of the potential is shown in
Fig. (\ref{sketch}).  The case $a=b=0$ corresponds to conventional F-
and D-term hybrid inflation. The case $a<0$ has been considered in the framework of F-term \cite{BasteroGil:2006cm, Boubekeur:2005zm}, D-term
\cite{Boubekeur:2005zm, Lin:2006xta, Lin:2007va} and $F_D$-term \cite{Garbrecht:2006az} hybrid inflation. The case $c=0$ has been considered in \cite{Kohri:2007gq, German:2001tz}.

In order to achieve the slow-roll condition, we require $V\simeq V_0$, giving
\begin{equation}
\frac{V^{\prime}}{V}=a\phi-b\phi^3+\frac{c}{\phi}
\end{equation}
\begin{equation}
\frac{V^{\prime\prime}}{V}=a-3b\phi^2-c/\phi^2 \equiv \eta
\end{equation}
Then Eq. (\ref{efolds}) has the analytic solution
\begin{equation}
\phi^2=\frac{(1+B)\sqrt{4bc+a^2}+(B-1)a}{2b(B-1)}
\end{equation}
where
\begin{equation}
B \equiv A e^{2\sqrt{4bc+a^2}N}
\end{equation}
and
\begin{equation}
A \equiv \frac{2b\phi^2_{end}+\sqrt{4bc+a^2}-a}{2b\phi^2_{end}-\sqrt{4bc+a^2}-a}.
\end{equation}
The spectral index is
\begin{equation}
n_s \simeq 1+2\eta.
\end{equation}
The curvature perturbation is
\begin{equation}
P_R=\frac{1}{12\pi^2}\frac{V_0}{\left(a\phi-b\phi^3+\frac{c}{\phi}\right)^2}.
\end{equation}

\section{Cosmic Strings}
\label{cosmicstring} 
\begin{figure}[htbp]
\begin{center}
\includegraphics[width=0.45\textwidth]{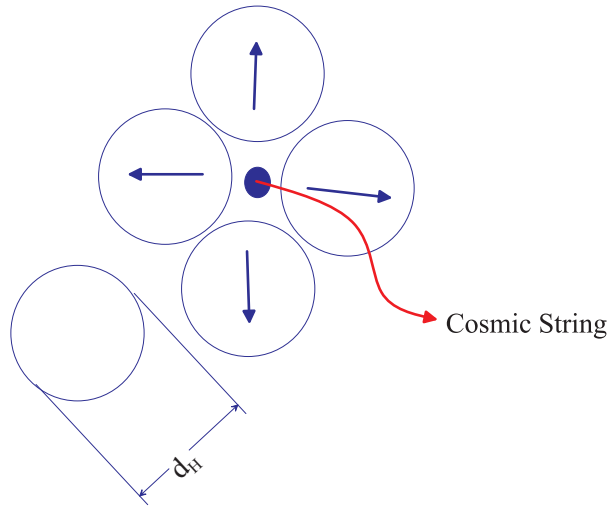}
\caption{The Universe can be divided into different parts by correlation length no larger than the particle horizon $d_H$. Each part has
different phase of the field represented by different directions of the arrow, result in cosmic string formation.}
\label{cs}
\end{center}
\end{figure}

In hybrid inflation, when the inflaton rolls
down its potential to the critical point, inflation ends via the
tachyonic instability of the waterfall field. The potential then
goes to its global minimum with the VEV of the waterfall field.
After hybrid inflation, there can be a $U(1)$ gauge symmetry broken
by the VEV of the waterfall field and thus cosmic strings (for
review, see \cite{Hindmarsh:1994re}) form via the Kibble mechanism
\cite{Kibble:1976sj}. The Kibble mechanism says that when the
symmetry breaking happens, different patches (at least smaller than
the particle horizon) of the Universe end up with different vacua.
At the boundaries between patches, topological defects (in our case,
cosmic string) form (Fig. (\ref{cs})). The original version of
Kibble mechanism describes the symmetry breaking when the Universe
cooled down from high temperature. Hybrid inflation is an
alternative to achieve the symmetry breaking.

The CMB anisotropy is analyzed by decomposing the fluctuation into
spherical harmonics given by \cite{Liddle:2000cg}
\begin{equation}
\frac{\Delta T(\theta, \phi)}{T}=\sum_{l,m}a_{lm}Y_{lm}(\theta,\phi),
\end{equation}
\begin{equation}
C_l \equiv \frac{1}{2l+1}\sum^{m=+l}_{m=-l}|a_{lm}|^2,
\end{equation}
where $Y_{lm}$'s are spherical harmonic functions and $C_l$ is the
angular power spectrum.
It is highly nontrivial to calculate the effect of cosmic string
network to CMB angular power spectrum, but roughly we can write
\cite{Endo:2003fr}
\begin{equation}
l(l+1)C^{str.}_l \propto (G\mu)^2
\end{equation}
where $G$ is Newton's constant and $\mu$ is the string tension. In
\cite{Bevis:2006mj} the value
\begin{equation}
G\mu=2.0\times 10^{-6}
\label{simu}
\end{equation}
was obtained using a field-theory simulation of the cosmic string
network when the cosmic string contribution to CMB angular spectrum
is assumed $100\%$ (at $l=10$).

As mentioned in Sec. \ref{1}, there is a $10\%$ upper bound for the
contribution of cosmic string network to the CMB angular power
spectrum from the observation data. Therefore,  the data puts an
upper bound on the string tension $\mu$.

\section{D-term Inflation}
\label{3}
The superpotential of D-term hybrid inflation is given by
\begin{equation}
W=\lambda S \Phi_+ \Phi_-
\end{equation}
where $S$ is the inflaton superfield, $\lambda$ is the
superpotential coupling, and $\Phi_\pm$ are chiral superfields
charged under the $U(1)_{FI}$ gauge symmetry responsible for the
Fayet-Iliopoulos term. The corresponding scalar potential is
\begin{equation}
V(S, \Phi_+, \Phi_-)=\lambda^2\left[|S|^2(|\Phi_+|^2+|\Phi_-|^2)+|\Phi_+|^2|\Phi_-|^2\right]+\frac{g^2}{2}\left(|\Phi_+|^2-|\Phi_-|^2+\xi\right)^2,
\end{equation}
where $\xi$ is the Fayet-Iliopoulos term and $g$ is the $U(1)_{FI}$
gauge coupling. A very small $g$ (far smaller than order $O(1)$) is
regarded as unnatural, because we do not know of any small (for
example $g<10^{-3}$) gauge couplings in particle physics. The true
vacuum of the potential is given by $|S|=|\Phi_+|=0$ and
$|\Phi_-|=\sqrt{\xi}$.  When $|S| > |S_c|= g \xi^{1/2}/\lambda$,
there is a local minimum occurred at $|\Phi_+|=|\Phi_-|=0$.
Therefore,  at tree level the potential is just a constant $V=g^2
\xi^2/2$. The 1-loop corrections to $V$ can be calculated using the
Coleman-Weinberg formula \cite{Coleman:1973jx}
\begin{equation}
\Delta V=\frac{1}{64\pi^2}\sum_i(-1)^F m^4_i\ln\frac{m^2_i}{\Lambda^2},
\end{equation}
with $m_i$ being the mass of a given particle, where the sum goes
over all particles with $F=0$ for bosons and $F=1$ for fermions and
$\Lambda$ is a renormalization scale. Thus the 1-loop potential is
given by (setting $\phi=\sqrt{2}\mbox{Re}(S)$)
\begin{equation}
V(S)=V_0\left(1+\frac{g^2}{4\pi^2}\ln\left(\frac{\phi^2}{\Lambda}\right)\right),
\end{equation}
where $V_0=g^2\xi^2/2$.

In the framework of supergravity, the D-term scalar potential is given by \cite{Lyth:1998xn}
\begin{equation}
V_D=\frac{1}{2}(\mbox{Re}f)^{-1}g^2(q_n K_n \Phi^n+\xi)^2,
\end{equation}
where $K$ is the K\"ahler potential and $K_n \equiv \partial K/
\partial \phi$. The gauge kinetic function $f$, which determines the
kinetic terms of the gauge and gaugino fields, is a holomorphic
function of all the complex scalar fields $\Phi_n$. One can choose
$f=1$ when the scalar fields are at the origin. Then, for example along
the $\Phi_1$ direction \cite{Lyth:1997ai},
\begin{equation}
1/f=1+\lambda_f \Phi_1^2/M_P^2+\cdots,
\end{equation}
where the linear term is assumed to be forbidden by a symmetry and
$\lambda$ is generically of order $O(1)$. In this paper, we consider
\begin{equation}
1/f=1+\alpha\frac{\phi^2}{M_P^2}-\beta\frac{\phi^4}{M_P^4}
\end{equation}
where $\phi$ is the inflaton field. Therefore, the scalar potential is
\begin{equation}
V=V_0\left(1+\alpha\frac{\phi^2}{M_P^2}\right)-\frac{V_0}{M_P^4}\beta \phi^4+V_0\frac{g^2}{4\pi^2}\ln\left(\frac{\phi}{\Lambda}\right),
\end{equation}
where $g$ is the $U(1)_{FI}$ gauge coupling and $\alpha$, $\beta$
are generically of order $O(1)$. If $\alpha,\beta$ are far from 1,
fine-tuning is needed.

After inflation the $U(1)_{FI}$ gauge symmetry is spontaneously
broken by the VEV of $\Phi_-$, cosmic strings form. The mass per
unit length of the string is given by
\begin{equation}
\mu=2 \pi \xi.
\label{dstring}
\end{equation}

From Eqs. (\ref{simu}, \ref{dstring}), we have $\xi^{1/2}=6.8 \times
10^{15} \mbox{GeV}$ for the cosmic string contribution to angular
spectrum being 100\%. We can calculate the upper bound on $\xi$ by
demanding the cosmic string contribution to the angular power
spectrum to be less than 10\% or 5\%.  We obtain $\xi^{1/2} \leq
4\times 10^{15} \mbox{GeV} =1.67\times 10^{-3} M_P$ for less than
10\% and $\xi^{1/2} \leq 1.34\times 10^{-3} M_P$ for less than $5\%$
contribution \cite{Bevis:2006mj}.

\begin{figure}[htbp]
\begin{center}
\includegraphics[width=0.45\textwidth]{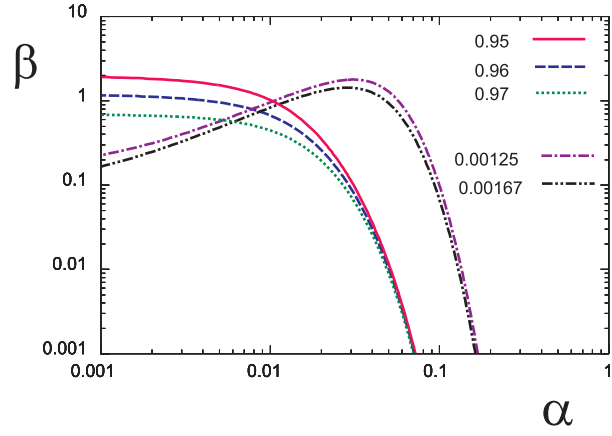}
\caption{$\phi_{end}=0.1$, $g=0.03$}
\label{1103}
\end{center}
\end{figure}

\begin{figure}[htbp]
\begin{center}
\includegraphics[width=0.45\textwidth]{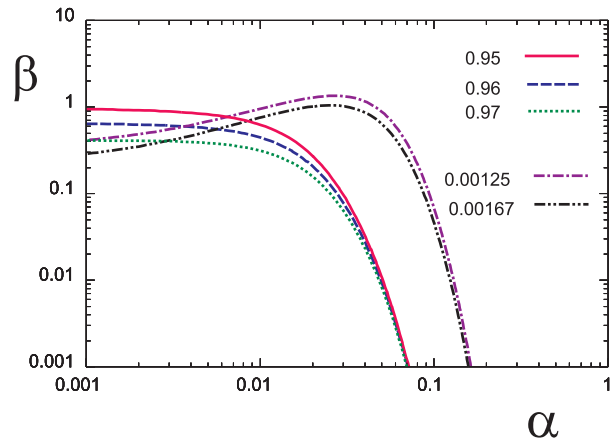}
\caption{$\phi_{end}=0.1$, $g=0.05$}
\label{1101}
\end{center}
\end{figure}

\begin{figure}[htbp]
\begin{center}
\includegraphics[width=0.45\textwidth]{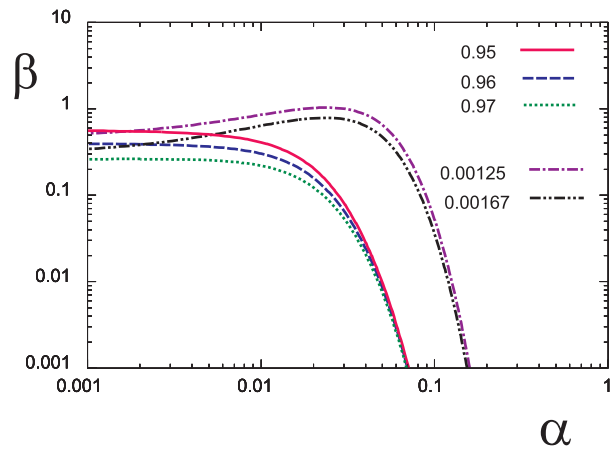}
\caption{$\phi_{end}=0.1$, $g=0.07$}
\label{1031}
\end{center}
\end{figure}

The results is shown in Figs. (\ref{1103}-\ref{1031}). We plot the
contours in $(\alpha,\beta)$ for
the spectral index $n_s=0.95, 0.96, 0.97$,
and for
$\xi^{1/2}=0.00167 M_P$ ($10\%$ cosmic string contribution) and
$\xi^{1/2}=0.00125 M_P$ (less than $5\%$ cosmic string
contribution). As we can see in the plots, the spectral index
$n_s=0.96$ can be achieved by mild tuning between
 the parameters $\alpha$ and $\beta$. Furthermore, if $g \lae
0.05$, cosmic string problem can be solved simultaneously at
$\alpha=O(10^{-2})$ and $\beta=O(1)$. Whereas $\lambda$ can be
calculated using $\phi_{end}\equiv
\sqrt{2}|S_c|=\sqrt{2}g\xi^{1/2}/\lambda$. For example, for
$\phi_{end}=0.1$, $g=0.05$, and $\xi^{1/2}=0.00125$, we obtain
$\lambda=8.83\times 10^{-4}$.

\section{F-term Inflation}
\label{4}
The superpotential of F-term Inflation is
\begin{equation}
W=\kappa S (\Phi_+\Phi_--M^2)
\end{equation}
where $S$ is the inflaton field, and $\Phi_+$, $\Phi_-$ are the
symmetry-breaking fields. The scalar potential is then
\begin{equation}
V=\kappa^2 \left|\Phi_+\Phi_--M^2\right|^2+\kappa^2|S|^2 \left(|\Phi_+|^2+|\Phi_-|^2|\right)+\frac{g^2}{2}\left(|\Phi_+|^2-|\Phi_-|^2\right)\,.
\end{equation}
Vanishing of the D-terms requires $|\Phi_-|=|\Phi_+|$. From the
potential, the true vacuum is given by $|S|=0$ and
$|\Phi_+|=|\Phi_-|=M$.  When $|S|>|S_c|=M$, a local minimum occurs
at $\Phi_+=\Phi_-=0$, and the corresponding potential at tree level
is just a constant $V_0=\kappa^2 M^4$. The 1-loop-corrected
potential is given by (setting $\phi=\sqrt{2}\mbox{Re}(S)$)
\begin{equation}
V=V_0 \left(1+\frac{\kappa^2}{8\pi^2}\ln\left(\frac{\phi}{\Lambda}\right)\right),
\end{equation}
where $V_0=\kappa^2 M^4$ and $\Lambda$ is a renormalization scale.
For F-term inflation, $\phi_{end} \equiv \sqrt{2}|S_c|=\sqrt{2}M$.

Consider a non-minimal K\"ahler potential
\cite{BasteroGil:2006cm, urRehman:2006hu}
\begin{equation}
K=|S|^2+|\Phi|^2+|\bar{\Phi}|^2+\kappa_S \frac{|S|^4}{4M_P^2}+\kappa_{S\Phi}\frac{|S|^2|\Phi|^2}{M_P^2}+\kappa_{S\bar{\Phi}}\frac{|S|^2|\bar{\Phi}|^2}{M_P^2}+\kappa_{SS}\frac{|S|^6}{6M_P^4}+\cdots.
\end{equation}
Here the couplings $\kappa_S$, $\kappa_{S\Phi}$,
  $\kappa_{S\bar{\Phi}}$ and $\kappa_{SS}$ are of order $O(1)$,
  because if they are not, the effective cutoff scale will not be $M_P$ but
  some unknown scales. The SUGRA F-term scalar potential is then given
  by
\begin{equation}
V_F=e^{\frac{K}{M^2_P}}\left[\left(W_m+\frac{W K_m}{M_P^2}\right)^\dag K^{m^\dag n} \left(W_n+\frac{WK_n}{M_P^2}\right)-\frac{3|W|^2}{M_P^2}\right],
\end{equation}
where $K^{m^\dag n}$ is the inverse matrix of
\begin{equation}
K_{m^\dag n}=\partial^2 K/\partial \phi^\dag_m \partial \phi_n.
\end{equation}

During inflation we have
\begin{equation}
V\simeq\kappa^2 M^4 \left(1-\kappa_S \frac{\phi^2}{2M_P^2}+\gamma_S\frac{\phi^4}{8M_P^4}+\frac{\kappa^2}{8\pi^2}\ln\left(\frac{\phi}{\Lambda}\right) \right)
\label{9}
\end{equation}
where
\begin{equation}
\gamma_S = (1-\frac{7\kappa_S}{2}+2\kappa^2_S-3\kappa_{SS}).
\end{equation}
Equation (\ref{9}) is of the form in Eq. (\ref{potential}).  We define
$\alpha \equiv -\kappa_S/2$ and $\beta \equiv -\gamma_S/8$. From
$\kappa_S$ we can see that $\alpha$ is about order $O(1)$. From $\gamma$ we
can see that the natural value of $\beta$ can be as large as something like
$O(10^2)$.  In \cite{BasteroGil:2006cm, urRehman:2006hu}, a
\emph{negative} quadratic correction is considered (corresponds to
negative $\alpha$). In this section, we consider \emph{positive}
quadratic correction and \emph{negative} quartic correction
(corresponds to $\alpha, \beta>0$).

Similar to D-term inflation, a cosmic string network forms after
F-term inflation with the string tension \cite{Hindmarsh:1994re,
Jeannerot:2005mc}
\begin{equation}
\mu=2\pi M^2 \theta(\beta),
\end{equation}
where $\theta(\beta) \sim 2.4 \ln(2/\beta)^{-1}$ and $\beta \simeq
(\kappa/g')^2$, where $g'$ is the appropriate gauge coupling, for
example, in the framework of GUT, $g' \simeq \sqrt{4\pi/25}$. The
constraint on $M$ is  $M \leq 2\times 10^{15} \mbox{GeV} = 8.3
\times 10^{-4} M_P $ under the requirement that cosmic string
contributes less than $10\%$ to the CMB angular power spectrum
\cite{Jeannerot:2005mc, Rocher:2004et}, while if we require the
contribution to be less than $5\%$ the corresponding constraint
becomes $M \leq 7\times 10^{-4} M_P$.

\begin{figure}[htbp]
\begin{center}
\includegraphics[width=0.45\textwidth]{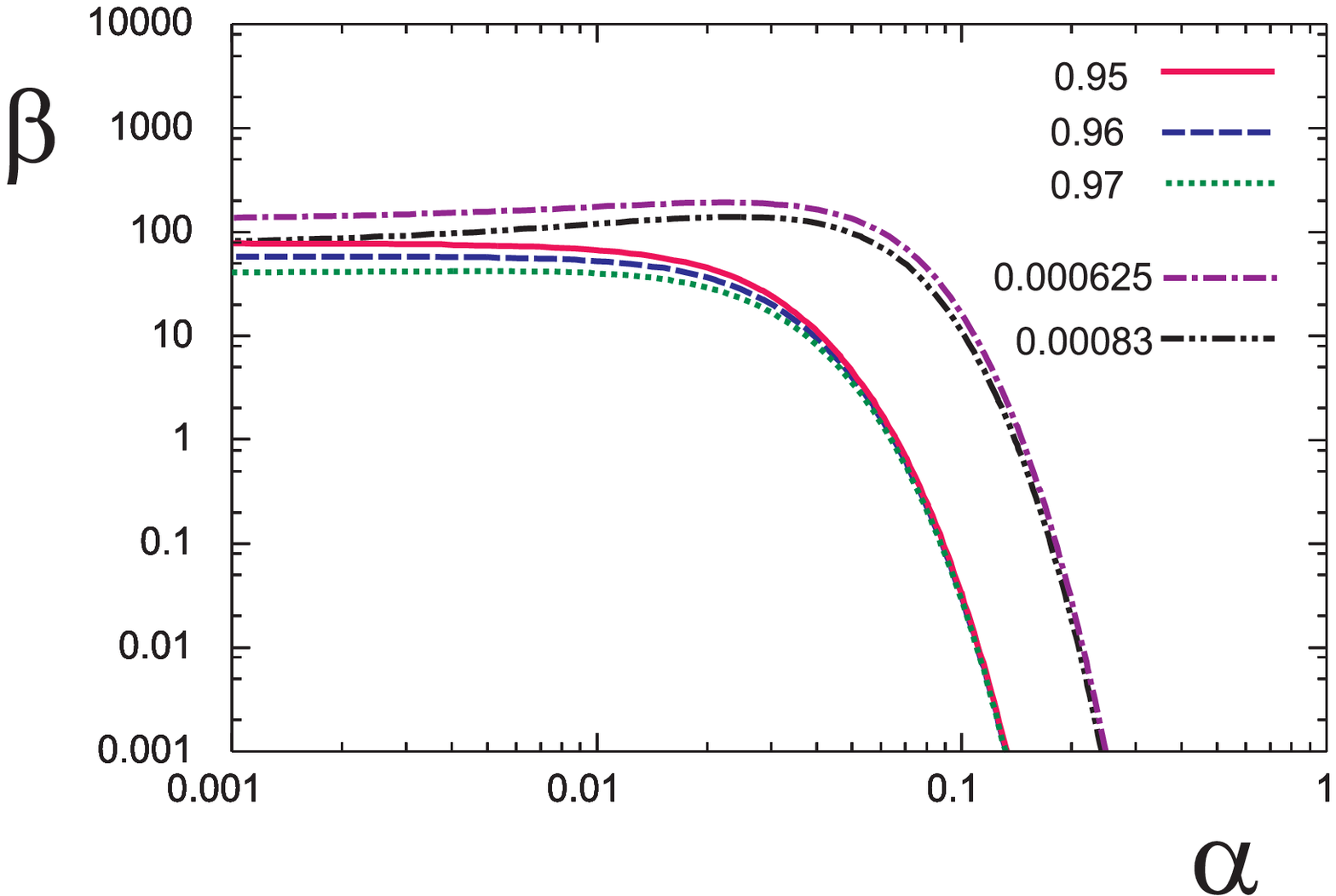}
\caption{$\kappa=0.01$}
\label{1107}
\end{center}
\end{figure}

\begin{figure}[htbp]
\begin{center}
\includegraphics[width=0.45\textwidth]{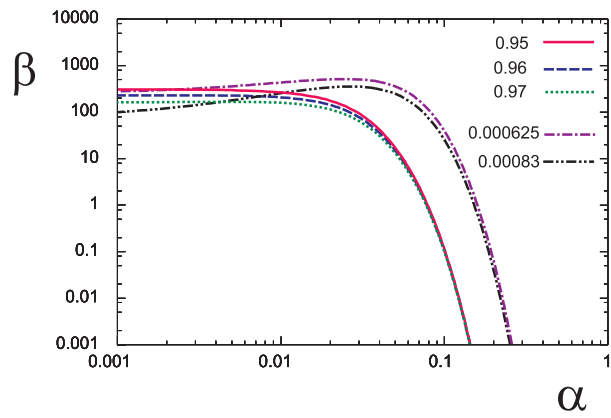}
\caption{$\kappa=0.005$}
\label{1105}
\end{center}
\end{figure}

\begin{figure}[htbp]
\begin{center}
\includegraphics[width=0.45\textwidth]{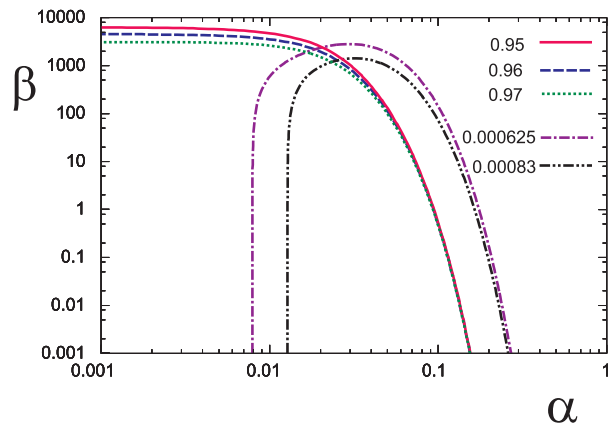}
\caption{$\kappa=0.001$}
\label{1106}
\end{center}
\end{figure}

The results are shown in Figs. (\ref{1107}-\ref{1106})
for various $\kappa$. We plot the
contours in $(\alpha,\beta)$
for the spectral index $n_s=0.95, 0.96, 0.97$, $M=8.3\times
10^{-4} M_P$ ($10\%$ cosmic string contribution) and $M=6.25\times
10^{-4} M_P$ (less than $5\%$ cosmic string contribution). As we can
see in the plots, $n_s=0.96$ can be achieved via mild tuning\footnote{In the framework of string theory, there is a
natural way to obtain small values of $\alpha$ \cite{Casas:1997uk,
  Casas:1999xj}.} between the
parameters $\alpha$ and $\beta$. Again, if $\kappa < 0.01$, cosmic string
problem can be solved simultaneously with $\alpha=10^{-1}-10^{-3}$ and
$\beta=10-10^3$.

\section{Conclusions}
\label{5}
In this paper we shown that the higher order corrections to the SUSY
hybrid inflation models can have very interesting results if we include quartic terms in addition to the quadratic term.
If the quadratic correction is positive and the quartic correction is
negative, a hilltop form for the inflaton potential is
possible, so that
the potential of such a hilltop form makes it possible to
have the spectral index reduced to $n_s=0.96$ after some mild tuning
of the coupling of the higher order terms. For D-term inflation, if $g
\lae 0.05$, cosmic string problem can be solved with
$\alpha=O(10^{-2})$ and $\beta=O(1)$. For F-term inflation, if $\kappa
< 0.01$, cosmic string problem can be solved with
$\alpha=10^{-1}-10^{-3}$ and $\beta=10-10^3$. Which needs more tuning
than the case of D-term inflation.

Lower bounds for $g$ and $\kappa$ are not crucial in this work. The
F- and D-term hybrid inflation works even for much lower $g$ or
$\kappa$, which means a lower scale $V_0$ and smaller $\epsilon$ (a
flatter potential). However, this more or less violates the spirit
of hybrid inflation which was introduced to prevent small couplings
to occur in the successful new inflation \cite{Linde:1981mu} or
chaotic inflation.

The model proposed in this paper can be applied not only restricted to
SUSY hybrid inflation. For example, the D3/D7 model
\cite{Dasgupta:2002ew} has an effective description as a D-term
inflation model and suffers the problem of producing too much cosmic
string contribution to the CMB. If higher order terms are generated in
this model, cosmic string problem can be evaded.

\section*{Acknowledgement}
This work was supported in part by the NSC under grant No. NSC
96-2628-M-007-002-MY3, by the NCTS, and by the Boost Program of
NTHU. We are grateful to D. H. Lyth, J. McDonald and Q. Shafi for
insightful comments. This work is also supported in parts by the WCU
program through the KOSEF funded by the MEST (R31-2008-000-10057-0).


\newpage
\begin{thebibliography}{99}

\bibitem{Starobinsky:1980te}
  A.~A.~Starobinsky,
  Phys.\ Lett.\  B {\bf 91}, 99 (1980).

\bibitem{Sato:1980yn}
  K.~Sato,
  Mon.\ Not.\ Roy.\ Astron.\ Soc.\  {\bf 195}, 467 (1981).


\bibitem{Guth:1980zm}
  A.~H.~Guth,
  Phys.\ Rev.\  D {\bf 23}, 347 (1981).

\bibitem{Lyth:1998xn}
  D.~H.~Lyth and A.~Riotto,
  Phys.\ Rept.\  {\bf 314}, 1 (1999)
  [arXiv:hep-ph/9807278].

\bibitem{Lyth:2007qh}
  D.~H.~Lyth,
  Lect.\ Notes Phys.\  {\bf 738}, 81 (2008)
  [arXiv:hep-th/0702128].

\bibitem{Linde:2007fr}
  A.~Linde,
  Lect.\ Notes Phys.\  {\bf 738}, 1 (2008)
  [arXiv:0705.0164 [hep-th]].


\bibitem{Komatsu:2008hk}
  E.~Komatsu {\it et al.}  [WMAP Collaboration],
  arXiv:0803.0547 [astro-ph].

\bibitem{Boubekeur:2005zm}
  L.~Boubekeur and D.~H.~Lyth,
  JCAP {\bf 0507}, 010 (2005)
  [arXiv:hep-ph/0502047].

\bibitem{Kohri:2007gq}
  K.~Kohri, C.~M.~Lin and D.~H.~Lyth,
  JCAP {\bf 0712}, 004 (2007)
  [arXiv:0707.3826 [hep-ph]].

\bibitem{Linde:1993cn}
  A.~D.~Linde,
  Phys.\ Rev.\  D {\bf 49}, 748 (1994)
  [arXiv:astro-ph/9307002].

\bibitem{Linde:1983gd}
  A.~D.~Linde,
  Phys.\ Lett.\  B {\bf 129}, 177 (1983).

\bibitem{Copeland:1994vg}
  E.~J.~Copeland, A.~R.~Liddle, D.~H.~Lyth, E.~D.~Stewart and D.~Wands,
  Phys.\ Rev.\  D {\bf 49}, 6410 (1994)
  [arXiv:astro-ph/9401011].

\bibitem{Dvali:1994ms}
  G.~R.~Dvali, Q.~Shafi and R.~K.~Schaefer,
  Phys.\ Rev.\ Lett.\  {\bf 73}, 1886 (1994)
  [arXiv:hep-ph/9406319].

\bibitem{Binetruy:1996xj}
  P.~Binetruy and G.~R.~Dvali,
  Phys.\ Lett.\  B {\bf 388}, 241 (1996)
  [arXiv:hep-ph/9606342].

\bibitem{Halyo:1996pp}
  E.~Halyo,
  Phys.\ Lett.\  B {\bf 387}, 43 (1996)
  [arXiv:hep-ph/9606423].

\bibitem{Riotto:1997wy}
  A.~Riotto,
  arXiv:hep-ph/9710329.

\bibitem{Lyth:1997pf}
  D.~H.~Lyth and A.~Riotto,
  Phys.\ Lett.\  B {\bf 412}, 28 (1997)
  [arXiv:hep-ph/9707273].

\bibitem{Pogosian:2003mz}
  L.~Pogosian, S.~H.~H.~Tye, I.~Wasserman and M.~Wyman,
  Phys.\ Rev.\  D {\bf 68}, 023506 (2003)
  [Erratum-ibid.\  D {\bf 73}, 089904 (2006)]
  [arXiv:hep-th/0304188].

\bibitem{Pogosian:2004ny}
  L.~Pogosian, M.~C.~Wyman and I.~Wasserman,
  JCAP {\bf 09}, 008 (2004)
  [arXiv:astro-ph/0403268].

\bibitem{Wyman:2005tu}
  M.~Wyman, L.~Pogosian and I.~Wasserman,
  Phys.\ Rev.\  D {\bf 72}, 023513 (2005)
  [Erratum-ibid.\  D {\bf 73}, 089905 (2006)]
  [arXiv:astro-ph/0503364].

\bibitem{Lin:2006xta}
  C.~M.~Lin and J.~McDonald,
  Phys.\ Rev.\  D {\bf 74}, 063510 (2006)
  [arXiv:hep-ph/0604245].


\bibitem{Endo:2003fr}
  M.~Endo, M.~Kawasaki and T.~Moroi,
  Phys.\ Lett.\  B {\bf 569}, 73 (2003)
  [arXiv:hep-ph/0304126].

\bibitem{Rocher:2004my}
  J.~Rocher and M.~Sakellariadou,
  Phys.\ Rev.\ Lett.\  {\bf 94}, 011303 (2005)
  [arXiv:hep-ph/0412143].



\bibitem{Bevis:2007gh}
  N.~Bevis, M.~Hindmarsh, M.~Kunz and J.~Urrestilla,
  Phys.\ Rev.\ Lett.\  {\bf 100}, 021301 (2008)
  [arXiv:astro-ph/0702223].

\bibitem{Battye:2006pk}
  R.~A.~Battye, B.~Garbrecht and A.~Moss,
  JCAP {\bf 0609}, 007 (2006)
  [arXiv:astro-ph/0607339].

\bibitem{Rocher:2004et}
  J.~Rocher and M.~Sakellariadou,
  JCAP {\bf 0503}, 004 (2005)
  [arXiv:hep-ph/0406120].


\bibitem{Dine:1995uk}
  M.~Dine, L.~Randall and S.~D.~Thomas,
  Phys.\ Rev.\ Lett.\  {\bf 75}, 398 (1995)
  [arXiv:hep-ph/9503303].


\bibitem{BasteroGil:2006cm}
  M.~Bastero-Gil, S.~F.~King and Q.~Shafi,
  Phys.\ Lett.\  B {\bf 651}, 345 (2007)
  [arXiv:hep-ph/0604198].


\bibitem{Lin:2007va}
  C.~M.~Lin and J.~McDonald,
  Phys.\ Rev.\  D {\bf 77}, 063529 (2008)
  [arXiv:0710.4273 [hep-ph]].

\bibitem{Garbrecht:2006az}
  B.~Garbrecht, C.~Pallis and A.~Pilaftsis,
  JHEP {\bf 0612}, 038 (2006)
  [arXiv:hep-ph/0605264].

\bibitem{German:2001tz}
  G.~German, G.~G.~Ross and S.~Sarkar,
  Nucl.\ Phys.\  B {\bf 608}, 423 (2001)
  [arXiv:hep-ph/0103243].

\bibitem{Hindmarsh:1994re}
  M.~B.~Hindmarsh and T.~W.~B.~Kibble,
  Rept.\ Prog.\ Phys.\  {\bf 58}, 477 (1995)
  [arXiv:hep-ph/9411342].


\bibitem{Kibble:1976sj}
  T.~W.~B.~Kibble,
  J.\ Phys.\ A  {\bf 9}, 1387 (1976).

\bibitem{Liddle:2000cg}
  A.~R.~Liddle and D.~H.~Lyth,


\bibitem{Bevis:2006mj}
  N.~Bevis, M.~Hindmarsh, M.~Kunz and J.~Urrestilla,
  Phys.\ Rev.\  D {\bf 75}, 065015 (2007)
  [arXiv:astro-ph/0605018].

\bibitem{Coleman:1973jx}
  S.~R.~Coleman and E.~J.~Weinberg,
  Phys.\ Rev.\  D {\bf 7}, 1888 (1973).

\bibitem{Lyth:1997ai}
  D.~H.~Lyth,
  Phys.\ Lett.\  B {\bf 419}, 57 (1998)
  [arXiv:hep-ph/9710347].


\bibitem{urRehman:2006hu}
  M.~ur Rehman, V.~N.~Senoguz and Q.~Shafi,
  Phys.\ Rev.\  D {\bf 75}, 043522 (2007)
  [arXiv:hep-ph/0612023].


\bibitem{Jeannerot:2005mc}
  R.~Jeannerot and M.~Postma,
  JHEP {\bf 0505}, 071 (2005)
  [arXiv:hep-ph/0503146].



\bibitem{Casas:1997uk}
  J.~A.~Casas and G.~B.~Gelmini,
  Phys.\ Lett.\  B {\bf 410}, 36 (1997)
  [arXiv:hep-ph/9706439].

\bibitem{Casas:1999xj}
  J.~A.~Casas, G.~B.~Gelmini and A.~Riotto,
  Phys.\ Lett.\  B {\bf 459}, 91 (1999)
  [arXiv:hep-ph/9903492].

\bibitem{Linde:1981mu}
  A.~D.~Linde,
  Phys.\ Lett.\  B {\bf 108}, 389 (1982).

\bibitem{Dasgupta:2002ew}
  K.~Dasgupta, C.~Herdeiro, S.~Hirano and R.~Kallosh,
  Phys.\ Rev.\  D {\bf 65}, 126002 (2002)
  [arXiv:hep-th/0203019].

\end{thebibliography}
\end{document}